\documentstyle[preprint,tighten,version2,aps]{revtex}

{\makeatletter%
\gdef\labeleqs#1{{%
\edef\@currentlabel{%
\ifappendixon\appletter\fi
\ifsecnumbers\ifnum\c@secnum>0
\arabic{secnum}.\fi\fi\arabic{equation}}%
\label{#1}%
}}%
}
\begin{document}
\draft
\preprint{IFUP-TH 57/95}
\begin{title}
Improved lattice operators: the case of the topological charge density.
\end{title}
\author{C. Christou$^a$, A. Di Giacomo$^b$, H. Panagopoulos$^a$, E. Vicari$^b$}
\begin{instit}
$^a$ Department of Natural Sciences, University of Cyprus, Nicosia

$^b$ Dipartimento di Fisica dell'Universit\`a and I.N.F.N., Pisa, Italy.
\end{instit}
\begin{abstract}
We analyze the properties of a class of improved
lattice topological charge density operators,
constructed by a smearing-like procedure.
By optimizing the choice of the parameters introduced in their definition,
we find operators having (i) a  better statistical behavior as estimators
of the topological charge density on the lattice, i.e. less noisy;
(ii)  a multiplicative renormalization much
closer to one; (iii) a large suppression of the perturbative tail
(and other unphysical mixings) in the corresponding lattice
topological susceptibility.
\end{abstract}



\section{Introduction.}
In QCD an important role is played by topological properties.
By the axial anomaly, matrix
elements or correlation functions involving the topological charge density
operator $q(x)$ can be related to relevant quantities of hadronic
phenomenology. We mention the topological susceptibility
$\chi$, which is
determinant in the explanation of the $U_A(1)$
problem~\cite{Witten},
and the on-shell nucleon matrix element of $q(x)$,
which can be related to the so-called
spin content of the nucleon~\cite{Carlitz}.

Lattice techniques represent our best source of non-perturbative
calculations, however investigating the topological properties of QCD on the
lattice is a non-trivial task. In a lattice theory the field is
defined on a discretized set and therefore the associated topological
properties are strictly trivial. One relies on the fact that the physical
continuum topological properties should be recovered in the continuum
limit.

{}From a field theoretical point of view, i.e. considering the lattice
as a regulator, difficulties may come from unphysical
divergences proportional to powers of the cut-off,
which must be removed and therefore
make the extraction of the physical signal hard.
In order to get reliable quantitative estimates of physical quantities,
one should control the unphysical cut-off dependent corrections
even when they disappear in the continuum limit,
given that numerical simulations
are performed at finite lattice spacings, i.e. at finite values of the
cut-off. Such corrections may be relevant, in that
the typical values of the bare coupling $g_0^2$ where
simulations are usually performed are actually
not small, but $g_0^2\simeq 1$, thus few terms in
perturbation theory are not always reliable.

Considering a lattice version of $q(x)$, $q_{_L}(x)$,
the classical continuum
limit must be in general corrected by including a renormalization
function. In pure QCD, where $q(x)$ is renormalization
group invariant, \cite{CDP}
\begin{equation}
q_{_L}(x)\;\rightarrow a^4 Z(g_0^2) q(x) + O(a^6)\;,
\label{renorm}
\end{equation}
where $Z(g_0^2)$ is a finite function of the bare
coupling $g_0^2$ going to one in the limit $g_0^2\rightarrow 0$,
but at $g_0^2\simeq 1$
it may be very different from one.
The finite renormalization of the widely used lattice operator~\cite{DiVecchia}
\begin{equation}
q_{_L}(x)=- {1\over 2^9\pi^2}
\sum^{\pm 4}_{\mu\nu\rho\sigma=\pm 1}
\epsilon_{\mu\nu\rho\sigma} {\rm Tr}
\left[ \Pi_{\mu\nu}\Pi_{\rho\sigma}\right]\;\,
\label{stop}
\end{equation}
($\Pi_{\mu\nu}(x)$ is the product of link variables
$U_\mu(x)$ around a $1\times 1$
plaquette) is quite nonnegligible:
for $SU(3)$ $Z(g_0^2=1)\simeq 0.18$~\cite{ACDDPV}.

The relation of the zero-momentum correlation
of two $q_{_L}(x)$ operators
\begin{equation}
\chi_{_L}\;=\; \sum_x \langle q_{_L}(x)q_{_L}(0) \rangle
\label{childef}
\end{equation}
with the topological susceptibility
$\chi$ is further complicated by an unphysical background term,
which eventually becomes dominant in the continuum limit.
(We recall that
the definition of $\chi$ requires also a prescription
to define the product of operators~\cite{Crewther}.)
Indeed
\begin{equation}
\chi_{_L}(g_0^2)\;=\;a^4 Z(g_0^2)^2\chi\,+\,M(g_0^2)\;.
\label{chileq}
\end{equation}
Neglecting terms $O(a^6)$,
the background term $M(g_0^2)$
can be written in terms of mixings with the unity operator
(so-called perturbative tail scaling as $\sim a^0$) and with the
trace of the energy-momentum (scaling as $\sim a^4$).
In the case of the operator (\ref{stop}) and for $SU(3)$,
$M(g_0^2)$ is already  dominant at $g_0^2\simeq 1$:
it is about 85\% of $\chi_{_L}$ at $g_0^2=1$~\cite{ACDGVp}.
As a consequence the uncertainty on $\chi$ can be hardly
made smaller than $\simeq 10\%$ by using the operator (\ref{stop})
and the heating method to evaluate $Z(g_0^2)$
and $M(g_0^2)$~\cite{Teperh,DV,ACDGV}.

Another problem, which has come up in some studies  concerning
the lattice determination of the on-shell proton matrix
element of $q(x)$~\cite{Gupta,ACDDPV},
is that the lattice operator (\ref{stop})
is very noisy, requiring very accurate statistics
and therefore expensive simulations in order to get a reasonable
uncertainty on the final result. In view of a full QCD
lattice calculation the search for a better estimator
appears a necessary step.

We study, within the field theoretical
approach, the possibility of improving the lattice
estimator of $q(x)$ with respect to all the problems listed
above, that is we look for local versions of $q(x)$ which
are less noisy, have a multiplicative renormalization
closer to one, and whose corresponding $\chi_{_L}$ is not
dominated by the unphysical backgroung signal $M(g_0^2)$
in the region $g_0^2\simeq 1$. (Any $\chi_{_L}$ defined
from a local $q_{_L}(x)$ will eventually be dominated by
its perturbative tail in the continuum limit. For the purpose of
evaluating $\chi$ it would suffice to have a small tail
at $g_0^2\simeq 1$,
which should be already in the scaling region.)

\section{Improved topological charge density operators.}
Inspired by the widely used smearing techniques, we consider the following
set of operators defined in terms of smeared links $V^{(i)}_\mu(x)$:
\begin{equation}
q_{_L}^{(i)}(x)=
- {1\over 2^9\pi^2}
\sum^{\pm 4}_{\mu\nu\rho\sigma=\pm 1}
\epsilon_{\mu\nu\rho\sigma} {\rm Tr}
\left[ \Pi^{(i)}_{\mu\nu}\,\Pi^{(i)}_{\rho\sigma}\right],
\label{impop}
\end{equation}
where $\Pi^{(i)}_{\mu\nu}$
is the product of smeared links $V^{(i)}_\mu(x)$  around a $1\times 1$
plaquette.
Such smeared links are constructed by the following procedure:
\begin{eqnarray}
V_\mu^{(0)}(x)&\equiv& U_\mu(x)\nonumber \\
\widehat{V}_\mu^{(i)}(x)&=& (1-c)V_\mu^{(i-1)}(x)+{c\over 6} \sum_{\pm\nu,
\nu\neq\mu}
V_\nu^{(i-1)}(x)
V_\mu ^{(i-1)}(x+\nu) V_\nu^{(i-1)}(x+\mu)^\dagger\nonumber \\
V_\mu^{(i)}(x)&=& {\widehat{V}_\mu^{(i)}(x)\over
\left[{1\over N}{\rm Tr}\,\widehat{V}^{(i)}_\mu(x)^\dagger
 \widehat{V}_\mu^{(i)}(x)\right]^{1/2} }
\label{imppr}
\end{eqnarray}
where $V_{-\nu}^{(i)}(x) = V^{(i)}_\nu(x-\nu)^\dagger$.
$V_\mu^{(i)}(x)$ and therefore $q_{_L}^{(i)}(x)$ depend on the parameter
$c$, which can be tuned to optimize the properties of $q_{_L}^{(i)}(x)$.
All these operators have the correct
classical continuum limit, i.e. for $a\rightarrow 0$,
$q_{_L}^{(i)}(x)\rightarrow a^4 q(x)$.

Notice that the size of $q_{_L}^{(i)}(x)$ increases with increasing the
integer parameter $i$. Nevertheless
$q_{_L}^{(i)}(x)$ can be still considered
as local operators when  keeping $i$ fixed while approaching the continuum
limit.
 Also, as we shall see,
by optimizing the choice of the parameter $c$,
a good improvement with respect to $q_{_L}^{(0)}(x)\equiv q_{_L}(x)$
is already achieved for small values of $i$.
For $SU(2)$ the procedure (\ref{imppr}) keeps the smeared
links $V_\mu^{(i)}(x)$ belonging to the $SU(2)$ group, and it is equivalent
to the smearing procedures proposed in Ref.~\cite{APE}.
For $N\geq 3$ the smeared links no longer
belong to the $SU(N)$ group.

The procedure
(\ref{imppr}) may be used to improve
any local operator involving link variables.
Smearing methods to improve lattice estimators
have been already widely employed in the study of long distance
correlations, such as large Wilson loops and hadron
source operators.

One often adopts an equivalent ``Schr\"odinger picture'' of smearing,
whereby lattice operators retain their original definition, while all
links in the configuration undergo transformation (\ref{imppr}). Full
consistency of this picture would then require that $V_\mu^{(i)}(x)$
be unitary. (As it stands, $V_\mu^{(i)}(x)$ is only unitary in the
case of $SU(2)$.) Projecting a matrix V onto $SU(N)$ amounts to
finding $X\in SU(N)$ which minimizes ${\rm Tr}\left((X^\dagger -
V^\dagger)(X-V)\right)$, or equivalently maximizes ${\rm Tr}\left(X^\dagger
V + V^\dagger X\right)$. The solution is given by:
\begin{equation}
X = {\rm i} \alpha V^{-1} + \left(V^\dagger V - \alpha^2 I\right)^{1/2}
V^{-1}
\label{unitary}
\end{equation}
where $\alpha$ is the real root of the equation:
\begin{equation}
\prod_i\left((d_i^2 - \alpha^2)^{1/2} + {\rm i} \alpha\right) = {\rm det}
V
\end{equation}
and $d_i\geq 0$ are the eigenvalues of $(V^\dagger V)^{1/2}$.
It can be verified that the lower loop results for $Z(g_0^2)$ and
the perturbative tail
$P(g_0^2)$ (see Sec.~\ref{pertan})
are not modified by rendering $V_\mu^{(i)}(x)$ unitary as
above.
It is worth mentioning at this point that abrupt cooling leads to exactly the
same unitary links $X$, for $c=1$. Indeed, cooling reassigns to each
link a new value, $X_\mu(x)$ in a way as to minimize the action,
i.e. maximize: ${\rm Tr}(X_\mu(x) V_\mu^{(i)}(x)^\dagger +
X_\mu(x)^\dagger V_\mu^{(i)}(x))$
at $c=1$, which coincides with Eq.(\ref{unitary}).

For $N\geq 3$, instead of projecting back onto the $SU(N)$ group
we propose last step of the procedure (\ref{imppr}), which is simpler
and should retain most of the advantages of the standard smearing procedure.

\section{Perturbative analysis.}
\label{pertan}
We have calculated $Z^{(1)}(g_0^2)$ to one loop for the once-smeared
operator $q_{_L}^{(1)}(x)$ with the Wilson action.
To carry out this calculation, $q_L^{(1)}(x)$
is expanded in a Taylor series
in the gauge field $A_\mu(x)$, where $U_\mu(x) = \exp(ig_0 A_\mu(x))$.
In Fig.~\ref{fig1} we show the three diagrams contributing to
$Z^{(1)}$. We find
\begin{eqnarray}
Z^{(1)}&=& 1 + z_1g_0^2 + O(g_0^4)\;,\nonumber \\
z_1&=&N\left[
{1\over 4N^2} -{1\over 8} -{1\over 2\pi^2}-I_0
+c\left(0.67789 - {0.24677\over N^2}\right)
+ c^2 \left(-0.48436 + {0.03991\over N^2}\right)\right] \;,
\label{mr}
\end{eqnarray}
where $I_0=0.15493$.
At $c=0$ we recover the non-smeared results~\cite{CDP}:
\begin{eqnarray}
Z&=& 1-0.5362 g_0^2+O(g_0^4) \;\;\;\;\;{\rm for} \;\;\;\; N=2\;,\nonumber \\
Z&=&1-0.9084 g_0^2+O(g_0^4)\;\;\;\;\;{\rm for} \;\;\;\; N=3\;,
\label{stopz}
\end{eqnarray}
which do not lead to a reliable estimate of $Z(g_0^2\simeq 1)$. As $c$
varies, the following extreme values of $Z$ are obtained:
\begin{eqnarray}
Z&=& 1-0.1360g_0^2+O(g_0^4) \;\;\;\;\; (c=0.6495) \;\;\;\;\; N=2\;,\nonumber\\
Z&=&1-0.2472g_0^2+O(g_0^4) \;\;\;\;\; (c=0.6774) \;\;\;\;\; N=3\;,
\label{min}
\end{eqnarray}
In both cases, $Z$ is quite close to unity for typical values of
$g_0^2$, making the one loop estimate more reliable.
It is noteworthy that the last step  in the smearing procedure (\ref{imppr})
turns out to be essential to make $Z$ approach one for $c\geq 0$.

For $q_{_L}^{(1)}(x)$, we have also calculated the lowest perturbative
contribution to the
mixing with the unity operator $P(g_0^2)$, which is the dominant part of the
background term $M(g_0^2)$ in the continuum limit.
The corresponding diagram is shown in Fig.~\ref{fig2}
and leads to the result:
\begin{eqnarray}
P(g_0^2) &=& g_0^6 {3 N(N^2-1)\over 128\pi^4} p(c)+ O(g_0^8)\;,\nonumber \\
p(c)&=&
0.002867 - 0.017685 c
+ 0.048665 c^2
- 0.075362 c^3\nonumber \\
&+& 0.068526 c^4
- 0.034433 c^5 + 0.007445 c^6\;.
\end{eqnarray}
The minimum of this everywhere-concave polynomial is
$p(c=0.872) = 1.4\times10^{-5}$.
Thus, for all $N$, the leading order of $P(g_0^2)$ diminishes by more than
two orders of magnitude compared to its non-smeared value ($c=0$).

In the presence of dynamical fermions one should
take into account the fact that, unlike pure gauge theory,
the topological charge density mixes under renormalization
with $\partial_\mu j_\mu^5$, where $j_\mu^5$ is the
singlet axial current.
The nonrenormalizability property of the anomaly in the ${\overline {\rm MS}}$
scheme means that the anomaly equation should take exactly the same form in
terms
of bare or renormalized quantities. However the renormalization of
$\partial_\mu j^5_\mu(x)$ and  $q(x)$
is nontrivial~\cite{Espriu}.
A renormalization group analysis leads to the following relation
valid for all matrix elements of a lattice version $q_{_L}(x)$ of $q(x)$
in the chiral limit~\cite{ADPV}:
\begin{equation}
\langle i2N_f q_{_L}\rangle\;=\;
Y(g_0^2)\,\langle R \rangle \;,
\label{ql}
\end{equation}
where $Y(g_0^2)$ is a finite function of $g_0^2$,
and
\begin{equation}
\langle R \rangle \;\equiv\;
\langle\; \partial_\mu j_\mu^5(x)_{R_{\overline {\rm MS}}}\;\rangle
\exp \int^0_{g(\mu)}
{\bar{\gamma}(\tilde{g})\over \beta_{_{\overline {\rm MS}}}(\tilde{g})} {\rm d}
\tilde{g}
\label{tt}
\end{equation}
is a renormalization group invariant quantity;
$\partial_\mu j_\mu^5(x)_{R_{\overline {\rm MS}}}$ indicates the operator
$\partial_\mu j_\mu^5(x)$ renormalized in the ${\overline {\rm MS}}$
scheme, and the function
$\bar{\gamma}(g)$ is related to the
anomalous dimension of the continuum operators $q(x)$, $\partial_\mu
j^5_\mu(x)$
in the ${\overline {\rm MS}}$ scheme:
$\bar{\gamma}(g)=
(1/16\pi^4) (3c_{_F}/2) N_f g^4+O\left(g^6\right)$.
Notice that
$\langle R \rangle $ is what can be naturally extracted also from experimental
data.

In perturbation theory one finds
$Y(g_0^2)=1+(z_1+y_1)g_0^2 + O(g_0^4)$, where $z_1$
is the coefficient of the $O(g_0^2)$ term of the finite renormalization
of $q_{_L}$ in the pure gauge theory (cfr. Eq.~(\ref{mr})), and
$y_1$ turns out to be a small number: $y_1=-0.0486$ for $N=3$
and $N_f=4$~\cite{ADPV}.

\section{Non-perturbative analysis by the heating method.}

Estimates of the multiplicative renormalizations of the
operators  $q_{_L}^{(i)}(x)$ and of the background term in the corresponding
$\chi_{_L}$  can be obtained using
the numerical heating method~\cite{Teperh,DV},
without any recourse to perturbation
theory.
This method relies on the idea that
the multiplicative renormalization $Z(g_0^2)$ and
the background term $M(g_0^2)$ is produced by short ranged
fluctuations at the scale of the cut-off $a$.
Therefore, when using a standard local algorithm
(for example Metropolis or heat bath)
to reach statistical equilibrium,
the modes contributing to $Z$ and $M$
should not suffer from critical slowing down, unlike
global quantities, such as the topological charge,
which should experience  a severe form of critical slowing~\cite{ACDGV}.

We applied the heating method to the operators
$q_{_L}^{(i)}(x)$ for $i=1,2$ and for a number
of values of $c$ in the region $0\leq  c \leq 1$.
We restricted our analysis to the $SU(2)$ pure gauge theory,
expecting no substantial differences for $N=3$.
The measurements were performed at $\beta=2.6$ ($g_0^2=1.5384...$),
which is a typical value for the $SU(2)$ simulations with the Wilson action.
The local updating was performed using the heat-bath algorithm.

An estimate of $Z$ can be obtained by heating
a configuration ${\cal C}_0$ which is
an approximate minimum of the lattice action and carries
a definite topological charge $Q_{L,0}$.
Such a configuration has been constructed by discretizing
an instanton solution in the singular gauge
\begin{equation}
A_\mu(x)\;=\; {\rho^2\over x^2+\rho^2} {i\over 2}
\left( s^\dagger_\mu s_\nu-s^\dagger_\mu s_\mu\right){x_\nu\over x^2}
\;,
\label{inst}
\end{equation}
where $s_4=1$ and $s_k=i\sigma_k$, and exponentiating
it to define link variables $U_\mu(x)=\exp iA_\mu(x)$.
Then a few cooling steps (about 5) were performed to make
the configuration smoother.
On a lattice $14^4$ and choosing $\rho=6$
we obtained an instanton-like configuration
carrying a topological charge $Q_{L,0}\simeq 0.96$ (all improved
operators we considered gave approximately
the same value for ${\cal C}_0$).

One then constructs
ensembles ${\cal C}_n$ of many independent
configurations obtained by heating ${\cal C}_0$ for the same number $n$
of updating steps, averaging $Q_{L}^{(i)}=\sum_x q_{_L}^{(i)}(x)$  over ${\cal
C}_n$
at fixed $n$. Let us define  $Q_{{L,n}}^{(i)}\equiv\langle
Q_L^{(i)}\rangle_{{\cal C}_n}$,
i.e. the average on the ensemble ${\cal C}_n$.
 Fluctuations of length $l\simeq a$ contributing
to $Z$ should rapidly thermalize, while the topological structure
of the initial configuration is left (approximately) unchanged
for a long time.
After a few heating steps where the short
ranged modes contributing to $Z$ get thermalized,
$Q_{{L,n}}$ should
show a plateau approximately at $Z\,Q_{{L,0}}$.
The estimates of $Z^{(i)}(\beta=2.6)$ from the plateaux observed in the heating
procedure  are reported in Table~\ref{tab1},
and should be compared with the value
$Z(\beta=2.6)=0.25(2)$ for the standard operator (\ref{stop})~\cite{ACDGV}.
The plateaux formed by the ratios $Q_{{L,n}}^{(i)}/Q_{{L,0}}^{(i)}$
starting from  $n\simeq 6$ are clearly observed in Fig.~\ref{zeta},
where data for $i=1,2$ and $c=0.8$ are plotted versus $n$.
Checks of the stability of the background topological structure
of the initial configuration were performed at $n=8,10,$
by cooling back the configurations (locally minimizing the action)
finding $Q_{L}\simeq Q_{{L,0}}$ after few cooling steps.

This analysis confirms the one-loop perturbative calculations,
that is the improved operators we considered have a multiplicative
renormalization closer to one than that of the initial operator $q_{_L}(x)$.
{}From  $Z(\beta=2.6)\simeq 0.25$ of $q_{_L}(x)$, we
pass, by roughly optimizing with respect to the parameter $c$, to
$Z^{(1)}(c=0.8,\beta=2.6)\simeq 0.57$ by one improving step,
and $Z^{(2)}(c=0.8,\beta=2.6)\simeq 0.75$
by two improving steps.
For larger $i$ we expect to get $Z^{(i)}$ closer and closer
to one, as also suggested from the results of the cooling method~\cite{CDPV}.
On the other hand, we should not forget that increasing
the number of improving steps the size of the operator
$q_{_L}^{(i)}(x)$ increases. One should find
a reasonable compromise taking into account the size of
the lattice one can afford in the simulations.

A comparison of the above results for $i=1$ with the one-loop
calculation (\ref{mr}) shows that the contribution of the
higher perturbative orders is still non-negligible,
but not so relevant as in the case of the operator without improving.

Another important property of the improved operators we can
infer from the heating method results is that they are much
less noisy than $q_{_L}(x)$ at fixed background. In other words,
in the Monte Carlo determinations of the matrix elements of $q(x)$
the contribution of short ranged
fluctuations to the error is largely suppressed.
A quantitative idea of this fact
may come from the quantity $e^{(i)}\equiv\Delta Z^{(i)}/Z^{(i)}$,
where $\Delta Z^{(i)}$
is the typical error of the data in the plateau during
the heating procedure described above.
We indeed found for $c\simeq 1.0$ and for an equal number of
measurements
\begin{eqnarray}
&&{e^{(0)}\over e^{(1)}}\;\simeq\; 6\;\,\nonumber \\
&&{e^{(0)}\over e^{(2)}}\;\simeq\; 15\;.
\label{eest}
\end{eqnarray}

An estimate of the background signal $M(g_0^2)$ can be obtained
by measuring $\chi_{_L,n}= \langle Q_{L}^2\rangle_{{\cal E}_n}/V$
on ensembles of configurations ${\cal E}_n$ constructed by
heating the flat configuration
for the same number $n$ of updating steps~\cite{DV,ACDGV}.
Measurements were performed on a $12^4$ lattice.
The plateau showed after few heating steps ($n\simeq 14$ in this case)
by the data of $\chi_{_L,n}^{(i)}$
should be placed approximately at the value of $M^{(i)}(g_0^2)$, since
no topological activity is detected there, i.e. the background is still flat
(this is checked by cooling back the heated configurations),
while the other modes contributing to $M(g_0^2)$ should be already
approximately thermalized (for a discussion of this issue
see \cite{ACDGV}).
The estimates of $M^{(i)}(\beta=2.6)$ from the plateaux observed
during heating are given in Table~\ref{tab2},
and should be compared with the value
$M(\beta=2.6)=2.10(5)\times 10^{-5}$
relative to  the standard operator (\ref{stop}).
In Fig.~\ref{tail} we plot $\chi^{(i)}_{_L,n}$ for $i=1,2$ and $c=1.0$
as a function of the heating step  $n$,
and compare with the corresponding data for the standard operator.
The expected plateaux are observed from $n\simeq 14$.

Notice the strong suppression of the background term in the improved operators.
For $c\simeq 1$ the reduction is about a factor 8 when performing one
improving step, and about a factor 30 by two improving steps.
For a larger number of improving steps, the suppression is
expected to be larger.

The suppression of the background term in Eq.~(\ref{chileq})
together with the relevant increase of $Z$ should
drastically change the relative weights of the contributions
to $\chi_{_L}$ in the relevant region
for Monte Carlo simulations.
By a standard Monte Carlo simulation
at $\beta=2.6$ on a $16^4$ lattice, we measured $\chi^{(i)}_{_L}$ for
$i=1,2$ and $c=0.6,0.8,1.0$.
We performed 15000 sweeps using an overrelaxed algorithm;
this sample size is already sufficient to show the better properties
of the operators $q_{_L}^{(i)}(x)$.
Data for $\chi_{_L}^{(i)}$ are given in Table~\ref{tab3}.
For comparison we also calculated $a^4\chi$
by cooling~\cite{Teperc}.

For the standard operator we found $\chi_{_L}=2.21(11)\times 10^{-5}$,
which, due to the large corresponding background term $M=2.10(5)\times
10^{-5}$,
does not allow one to determine  $a^4\chi$ at this value of $\beta$.
Instead the improved operators $q_{_L}^{(i)}(x)$ provide,
using Eq.~(\ref{chileq}), reliable  estimates of $a^4\chi$ having
about 10\% of uncertainty, which are consistent with each other
and are also consistent with the determination from cooling:
$\chi_{\rm cool}=1.3(2)\times 10^{-5}$,  although the latter
seems to be systematically lower. This fact may be explained taking
into account that $Q_L=\sum_x q_{_L}(x)$, which is used to estimate the
topological charge of cooled
configurations, underestimates the topological charge content
(for the lattice size we are working with), as we found out explicitly
when we constructed an instanton configuration on the lattice.

The determinations of $Z$ and $M$ should not be subject to relevant
finite size scaling effects (as explictly checked in Ref.~\cite{ACDGV}),
since they have their origin in short ranged
fluctuations. Thus finite size corrections to our estimates of $Z$ and $M$
should
be negligible.
Larger finite size effects are expected on the topological modes, as can
be argued from numerical studies available in the literature.
For this reason the measurement of $\chi_{_L}$, which receives contributions
also from topological modes, was performed on a larger lattice.
We should say that we did not perform a complete analysis of the finite size
corrections to $\chi_{_L}$, since our purpose was just to show the better
behavior of the improved operators $q_{_L}^{(i)}(x)$ and not the determination
of $\chi$ for the $SU(2)$ gauge model. So we limited ourselves to a numerical
study
not requiring a super-computer.

If the improvement for $SU(3)$ is similar to that achieved
for $SU(2)$, using the optimal operator for $i=2$ at $g_0^2=1$
the unphysical term in Eq.~(\ref{chileq})
is expected to become a small part of the total signal,
allowing a precise determination of $\chi$ by the field
theoretical method.

\section{Conclusions.}
We have analyzed the properties of a class of improved
lattice topological charge density operators constructed
by a smearing-like procedure.
Such operators look promising for the lattice
calculation of the on-shell proton matrix element of
the topological charge density operator
in full QCD, which is related to the so-called
proton spin content. Indeed their use should overcome the
difficulty due to the large noise observed in  preliminary
quenched studies~\cite{Gupta,ACDDPV}, and
they have a multiplicative renormalization much
closer to one.

Improved operators are also expected to provide a much better
determination
of the topological susceptibility by the field theoretical method
in the $SU(3)$ gauge theory, by strongly reducing the unphysical background
term
while enhancing the term containing $\chi$
with larger values of the multiplicative
renormalization. This should  allow a precise and independent
check of the alternative
cooling method determinations (see e.g. Refs.~\cite{Teperc,Teperc2}), whose
systematic errors are not completely
controlled.
Furthermore the improved operators may
also open the road to a more reliable lattice investigation of the
behavior of the topological susceptibility
at the deconfinement transition, where cooling
does not give satisfactory results~\cite{DMP}.

The smearing-like procedure (\ref{imppr}) may be used to improve
any local operator involving link variables, and a renormalization
study would again be called for in all cases. We hope to return to
this issue in the future.



\figure{One-loop diagrams contributing to the multiplicative renormalization
of $q_{_L}^{(i)}(x)$.
\label{fig1}}

\figure{Diagram contributing to the lowest order term of the perturbative
tail.
\label{fig2}}

\figure{$Q_{{L,n}}/Q_{{L,0}}$ versus $n$ in the heating
procedure of an instanton configuration.
Data for $i=1,2$ at $c=0.8$ are shown. For comparison the full
line represents the estimate of $Z$ for the standard operator (\ref{stop}).
\label{zeta}}

\figure{$\chi_{_L,n}$ versus $n$ in the heating procedure
of the flat configuration.
Data for the standard operator and improved operators for
$i=1,2$ at $c=1.0$ are shown.
\label{tail}}


\begin{table}
\caption{We present $Z^{(i)}(\beta=2.6)$ for $i=1,2$ and
various values of $c$, as obtained by the heating method
for a number of $\simeq 750$ trajectories.
The errors diplayed include both a statistical error
(determined by the typical errors of data in the plateau)
and a systematic error related to the stability of the
background configuration.
The numbers in this Table should be compared with the value
$Z(\beta=2.6)=0.25(2)$ for the standard operator (\ref{stop})~\cite{ACDGV}.
}
\label{tab1}
\begin{tabular}{rr@{}lr@{}lr@{}l}
\multicolumn{1}{r}{$i$}&
\multicolumn{2}{c}{$c=0.6$}&
\multicolumn{2}{c}{$c=0.8$}&
\multicolumn{2}{c}{$c=1.0$}\\
\tableline \hline
1 & 0&.52(2) & 0&.57(2) & 0&.54(2) \\
2 & 0&.68(2) & 0&.75(2) & 0&.68(2) \\
\end{tabular}
\end{table}

\begin{table}
\caption{We present $M^{(i)}(\beta=2.6)$ for $i=1,2$ and
various values of $c$, as obtained by the heating method
(1000 trajectories), i.e. from the the value of $\chi_{_L}$ at the plateau,
observed around $n\simeq 16$ for all operators considered.
The numbers in this Table should be compared with the value
$M(\beta=2.6)=2.10(5)\times 10^{-5}$ for the standard operator (\ref{stop}).
}
\label{tab2}
\begin{tabular}{rr@{}lr@{}lr@{}l}
\multicolumn{1}{r}{$i$}&
\multicolumn{2}{c}{$c=0.6$}&
\multicolumn{2}{c}{$c=0.8$}&
\multicolumn{2}{c}{$c=1.0$}\\
\tableline \hline
1 & 0&.60(2)$\times 10^{-5}$& 0&.37(2)$\times 10^{-5}$ &0&.27(2)$\times
10^{-5}$ \\
2 & 0&.23(2)$\times 10^{-5}$& 0&.13(2)$\times 10^{-5}$ &0&.07(2)$\times
10^{-5}$ \\
\end{tabular}
\end{table}

\begin{table}
\caption{We present $\chi_{_L}^{(i)}(\beta=2.6)$ for $i=1,2$ and
various values of $c$, as obtained by  a standard Monte Carlo
simulation on a $16^4$ lattice
(15000 sweeps using an overrelaxed algorithm).
We also give the corresponding values of $a^4\chi$ as obtained
from Eq.~(\ref{chileq}). Data for $a^4\chi$ must be compared with the
cooling result: $\chi_{\rm cool}=1.3(2)\times 10^{-5}$.
}
\label{tab3}
\begin{tabular}{lrr@{}lr@{}lr@{}l}
\multicolumn{1}{r}{$$}&
\multicolumn{1}{r}{$i$}&
\multicolumn{2}{c}{$c=0.6$}&
\multicolumn{2}{c}{$c=0.8$}&
\multicolumn{2}{c}{$c=1.0$}\\
\tableline \hline
$\chi_{_L}$& 1 & 1&.02(7)$\times 10^{-5}$ & 0&.89(7)$\times 10^{-5}$&
0&.74(7)$\times 10^{-5}$ \\
           & 2 & 0&.93(7)$\times 10^{-5}$ & 0&.92(7)$\times 10^{-5}$&
0&.71(6)$\times 10^{-5}$ \\
\hline
$a^4\chi=(\chi_{_L}-M)/Z^2$
           & 1 & 1&.5(3)$\times 10^{-5}$  & 1&.6(3)$\times 10^{-5}$&
1&.6(3)$\times 10^{-5}$ \\
           & 2 & 1&.5(2)$\times 10^{-5}$  & 1&.4(2)$\times 10^{-5}$&
1&.4(2)$\times 10^{-5}$ \\
\end{tabular}
\end{table}

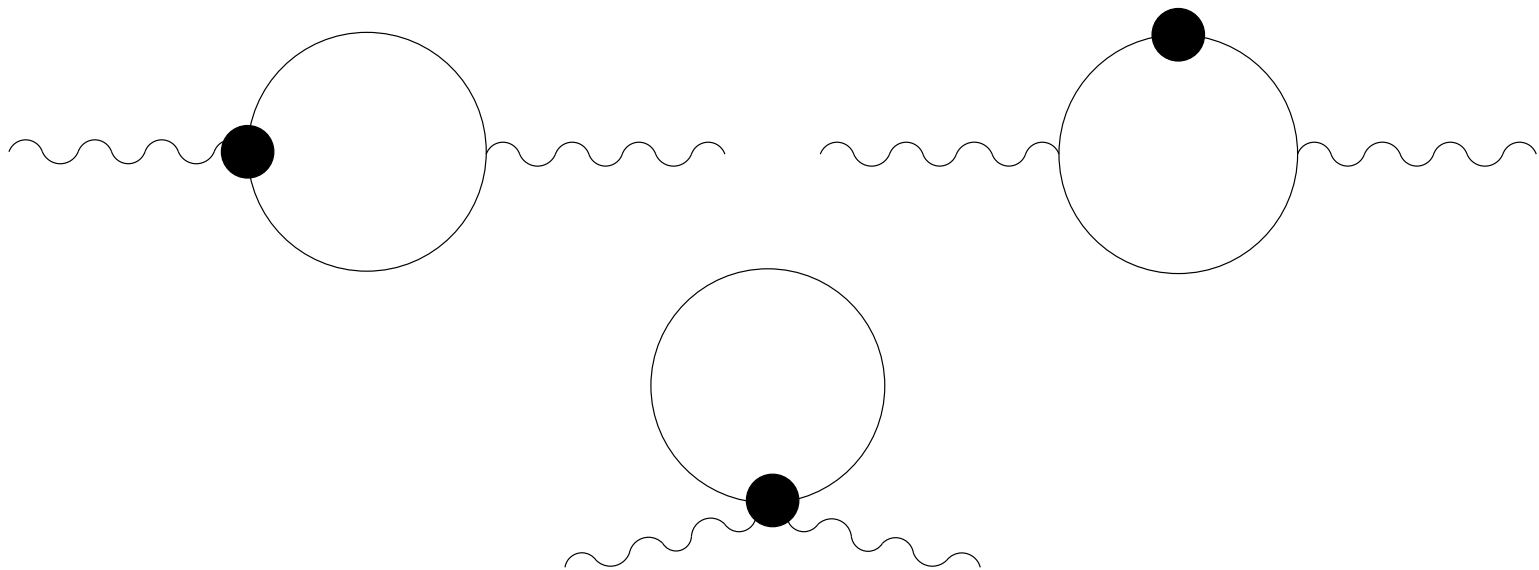
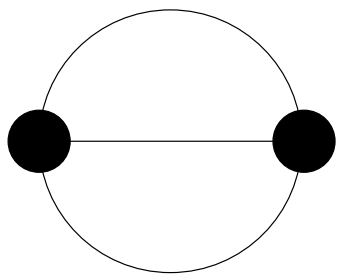
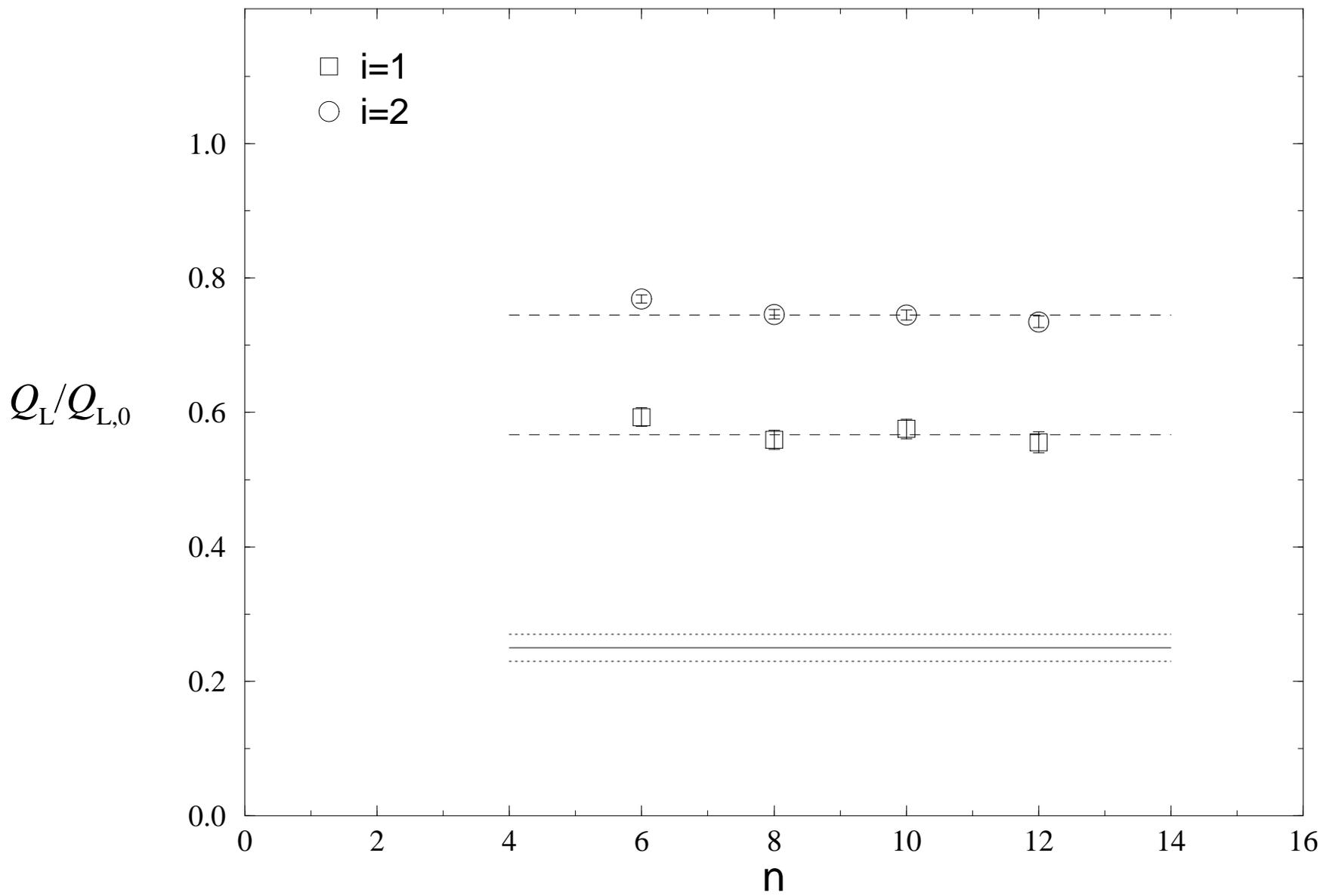
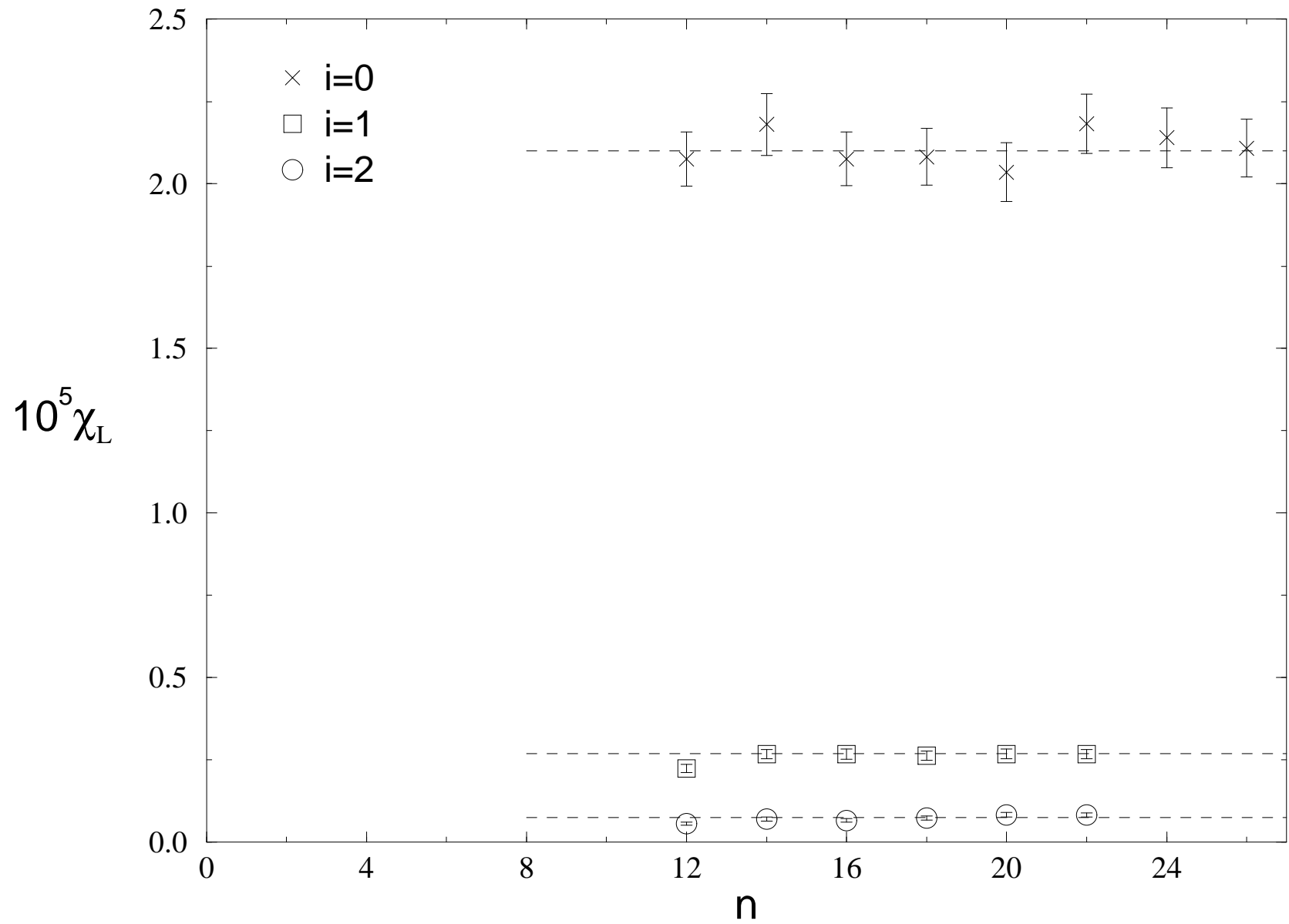
\end{document}